# Core/Shell CdSe/CdS Bone-Shaped Nanocrystals with a Thick and Anisotropic Shell as Optical Emitters

*Andrea Castelli, Balaji Dhanabalan, Anatolii Polovitsyn, Vincenzo Caligiuri, Francesco Di Stasio, Alice Scarpellini, Rosaria Brescia, Milan Palei, Beatriz Martin-Garcia, Mirko Prato, Liberato Manna, Iwan Moreels, Roman Krahne\*, and Milena P. Arciniegas\**

Dr. A. Castelli, B. Dhanabalan, Dr. V. Caligiuri, Dr. F. Di Stasio, A. Scarpelini, Dr. R. Brescia, Dr. M. Palei, Dr. B. Martin-Garcia, Dr. M. Prato, Prof. L. Manna, Dr. R. Krahne, and Dr. M.P. Arciniegas.
Istituto Italiano di Tecnologia, Via Morego 30, 16163 Genova, Italy.
*E-mails: Roman.Krahne@iit.it and Milena.Arciniegas@iit.it

Dr. A. Polovitsyn and Prof. Iwan Moreels.
Department of Chemistry, Ghent University, Krijgslaan 281-S3, 9000 Gent, Belgium

B. Dhanabalan.
Dipartimento di Chimica e Chimica Industriale, Università degli Studi di Genova, Via Dodecaneso, 31, 16146, Genova, Italy.

Keywords: Core/shell nanocrystals, branched heterostructures, seeded-growth synthesis, dual emission, ASE and random lasing.

Colloidal core/shell nanocrystals are key materials for optoelectronics, enabling control over essential properties via precise engineering of the shape, thickness, and crystal lattice structure of their shell. Here, we apply the growth protocol for CdS branched nanocrystals on CdSe nanoplatelet seeds and obtain bone-shaped heterostructures with a highly anisotropic shell. Surprisingly, the nanoplatelets withstand the high growth temperature of 350 °C and we obtain structures with a CdSe nanoplatelet core that is overcoated by a shell of cubic CdS, on top of which tetrahedral CdS structures with hexagonal lattice are formed. These complex core/shell nanocrystals show a band-edge emission around 657 nm with a photoluminescence quantum yield of ca. 42 % in solution, which is also retained in thin films. Interestingly, the nanocrystals manifest simultaneous red and green emission, and the relatively long wavelength of the green emission indicates charge recombination at the cubic/hexagonal interface of the CdS shell. The nanocrystal films show amplified spontaneous emission, random lasing, and distributed feedback lasing when the material is deposited on suitable gratings. Our work stimulates the design and fabrication of more exotic core/shell heterostructures where charge carrier delocalization, dipole moment, and other optical and electrical properties can be engineered.

Colloidal nanocrystals (NCs) are highly attractive as optoelectronic materials for a wide range of applications.[1] Controlling their shape and size is paramount to tune their optoelectronic properties and to achieve high photoluminescence quantum yield (PLQY) along with narrow and tuneable emission.[2] Through different synthetic strategies there is now access to a large variety of structures, from single material quantum dots[3], nanorods[4] and nanoplatelets (NPLs)[2a, 5] to core/shell systems that provide increased stability and photoluminescence efficiency,[6] multifunctional particles that combine metallic and semiconductor portions,[7] and highly sophisticated morphologies, such as branched[8] and hollow NCs.[9] This wealth in structural and material design translates into a broad range of applications in light emission, energy conversion, catalysis, and medicine.[1b, 10]

Core/shell nanocrystals are particularly advantageous for light emission, in particular those with a CdSe core that are epitaxially overgrown with a shell made of higher band-gap materials, such as ZnS and CdS. This is due to their efficient defect passivation,[6a, 11] the possibility to tailor wavefunction overlap and oscillator strength,[12] piezoelectric effects that stem from strain induced at the interface between the core and the shell,[13] and the possibility to control the self-assembly and polarization via an anisotropic nanocrystal architecture.[14] The choice of shell material allows to tailor the band alignment in the heterostructure from type I (CdSe/ZnS)[11a] to quasi type II (CdSe/CdS),[6a] and fully type II (CdSe/CdTe),[15] which enables to tune the wavefunction overlap and reduce Auger recombination.[16] Auger recombination could also be significantly decreased by employing alloyed shells,[17] thick (giant) shells,[18] and gradient shell growth.[19]

Several successful strategies have been applied to fabricate anisotropic heterostructures from dot- and rod-shaped cores, leading to elongated and branched NC shapes, such as dot-in-rods[14a, 20], rod-in-rods,[13, 20] tetrapods,[8, 21] and octapods.[22] Concerning colloidal NPLs as cores, which represent a particularly interesting class of colloidal emitters due to their sharp emission line width and large oscillator strength, the synthesis of heterostructures with more elaborate shapes is still in its infancy.[5b, 23] The challenge for branched NC heterostructures with CdSe NPLs as core material resides in the high temperatures of the shell growth processes, which can cause structural damage to the NPLs.[24]

In this work, we fabricate CdSe/CdS NCs by applying the protocol for the CdS shell growth of the octapod synthesis starting from CdSe NPLs seeds.[5f, 22] We obtain core/shell heterostructures with the NPLs as cores that are coated by a double CdS shell: a first layer, surrounding the whole NPL core, characterized by a cubic crystal structure (sphalerite-type), and an outer branched shell with a hexagonal crystal structure (wurtzite-type), composed of two tetrahedral-shaped protrusions. Despite this very complex architecture, that we name NPLs in tetrahedron dimers (NPL-i-TDs), our synthesis yields nearly monodisperse NCs whose shape resembles that of a bone. The NPL-i-TDs manifest emission in the deep red spectral region (at 660 nm) with a linewidth of around 30 nm, show a stable PLQY of 42 %, and an amplified spontaneous emission (ASE) threshold of 270 μJ/cm$^2$. Furthermore, we observe a second emission peak at 544 nm under high pump fluence that originates from shell transitions. By combining our optical data with ultraviolet photoelectron spectroscopy (UPS) measurements, we draw a flat band diagram of these heterostructures that elucidates the origin of the red and green emission. Furthermore, the NPL-i-TDs show single mode lasing when deposited on distributed feedback resonators with a lasing threshold below 300 μJ/cm$^2$, and random lasing from NPL-i-TD drop-cast films with a threshold around 50 μJ/cm$^2$.

**Results**

The synthesis of CdSe NPLs is well established for NCs with 2-8 monolayer (ML) thickness that emit at defined wavelengths in the visible range.[2a, 5a, 5e] Additional control on the lateral size of the NPLs enables a fine tuning of the emission wavelength around these values.[5f] Here, we synthesized CdSe NPLs with 4.5 ML thickness, length of 30 ± 5 nm and width of 5 ± 1 nm, following a synthetic protocol reported by our group.[5f] More details of the NPLs are provided in Figure S1, including representatives bright-field transmission electron microscopy (BF-TEM) images and the X-ray diffraction (XRD) pattern, which shows their cubic crystal structure (sphalerite-type, see Figure S1d). The 4.5 ML thickness results in a strong quantization of the electronic states along the [001] direction, and leads to a sharp PL emission peak centred at 510 nm with a full width at half maximum (FWHM) of ca. 10 nm, and absorbance peaks located at 480 nm and 509 nm (see spectra in Figure S2), in agreement with previous reports.[5f] We used these NPLs as seeds in a second reaction where a CdS shell was grown at high temperature (350 °C), following our established protocol for branched NCs with minor modifications (see Experimental Section).[22]

Interestingly, this synthesis approach produced monodispersed particles with an unusual anisotropic shape, as shown in the high-angle annular dark field-scanning TEM (HAADF-STEM) image in **Figure 1a** and in the BF-TEM image in Figure S3a. The TEM projection of the NC shape features a reduction in width towards the centre, from ca. 20 nm ($p$) to 10 nm ($w$) as shown in detail for a single NPLs-i-TDs in Figure S3b, giving rise to bone-shaped particles. The length, $l$, of the synthesized NCs is 30 ± 3 nm, similar to those of the initial NPLs, which can be linked to the slower shell growth along the $[100]_{CdSe}$ direction (Figure S3b and Table S1) in the conditions used. The collected XRD patterns evidence a series of peaks that can be associated with cubic and hexagonal CdS polymorphs, while the signal of cubic CdSe seeds is not observed (Figure S3c). High-resolution TEM (HRTEM) analysis confirms the presence of a cubic CdS phase in the central region (red-framed panels in Figure 1b) of the heterostructures and the hexagonal crystal structure of the CdS pods (yellow-framed panels in Figure 1b). Assuming the cubic CdS shell grows epitaxially on the CdSe NPLs, the HRTEM-deduced relative orientation of the two structures (cubic and hexagonal CdS) is compatible with the epitaxial orientation reported for the CdSe(core)/CdS(pods) octapods,[22] i.e. $CdSe_{cub}(111)$//

CdS$_{cub}$(111)//CdS$_{hex}$(0001) and CdSe$_{cub}$[2-1-1]//CdS$_{cub}$[2-1-1]//CdS$_{hex}$ [10-10] (see details in Fig. S4). The presence of the CdSe NPL cores cannot be demonstrated by HRTEM, but it can be identified in HAADF-STEM imaging (see Figure 1c) due to the sensitivity of the latter technique to atomic number (Z) contrast. In these images, the higher intensity in thin stripes along the main axis of the NCs (indicated with a green arrow in Figure 1c) can be associated to the NPLs seeds, due to the presence of the Se atoms with higher Z. More images that evidence the presence of the CdSe cores in such structures are provided in Figure S5.

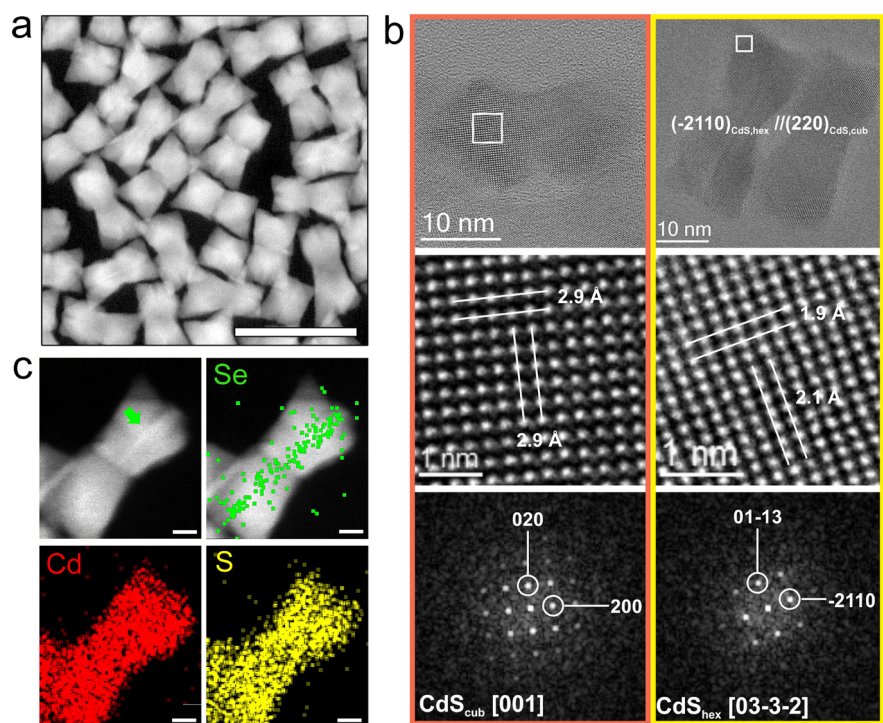

**Figure 1. a,** HAADF-STEM image of the NPL-i-TDs evidencing their monodispersity in shape and size. Scale bar: 50 nm. **b,** HRTEM analysis of the crystal structure of NPL-i-TDs: overall image of one particle, with corresponding zoom in the central region (red-framed panels) and in a pod (yellow-framed panels) followed by fast Fourier transforms (FFTs) demonstrating the cubic phase, compatible with CdS (ICSD 81925) and the hexagonal CdS (ICSD 154186), respectively. **c,** STEM-EDS compositional mapping of the distribution of Se, Cd, and S in a single particle, confirming the presence of Se from the cores in the brighter area in the centre of the NCs, while the rest of the structure is composed of CdS. Scale bars: 5 nm.

The qualitative indication provided by HAADF-STEM images on the location of CdSe cores in the heterostructures is confirmed by STEM-energy-dispersive X-ray spectroscopy (EDS) mapping (Figure 1c and S6), which shows that Se is localized in a thin region at the centre of the NCs, matching the thickness of the initial NPLs. The acquired information about the composition of the NPLs-i-TDs allows to outline their growth process as follows: when the CdSe NPLs are injected in the reaction mixtures, they serve as heterogeneous nucleation points for the epitaxial growth of the CdS shell. The lack of {111} facets in a NPL discourages the direct formation of an hexagonal CdS shell, and thus promoting the formation of first a cubic CdS shell. While this shell growth in thickness around the CdSe core, {111} facets might become available at the corners of the now CdSe core/cubic CdS shell structure, which provide suitable locations for the growth of a more stable hexagonal CdS shell that in a later stage crystallizes in pods, as it has been previously described by our group for CdSe/CdS branched NCs.[25]

The complex geometry of the synthesized heterostructures has been derived by HAADF-STEM tomography. Few chosen projections of a selected nanoparticle and the three-dimensional reconstructed volume are shown in Figure 2, as well as in the Movies S1 and S2. This analysis allows to clearly reveal the presence of two tetrahedra protruding from the central platelet. The arrangement of the tips of the tetrahedra agrees with the faster growth of the hexagonal CdS shell along four of the <111> directions of the cubic CdS. In most particles, the two tetrahedra are aligned in the same direction (see Figure 2b,c), indicating the preferential growth along either the [0001] or [000-1] hexagonal CdS direction: the two directions cannot be distinguished unless a specific study of the polarity is done.[26]

WILEY-VCH

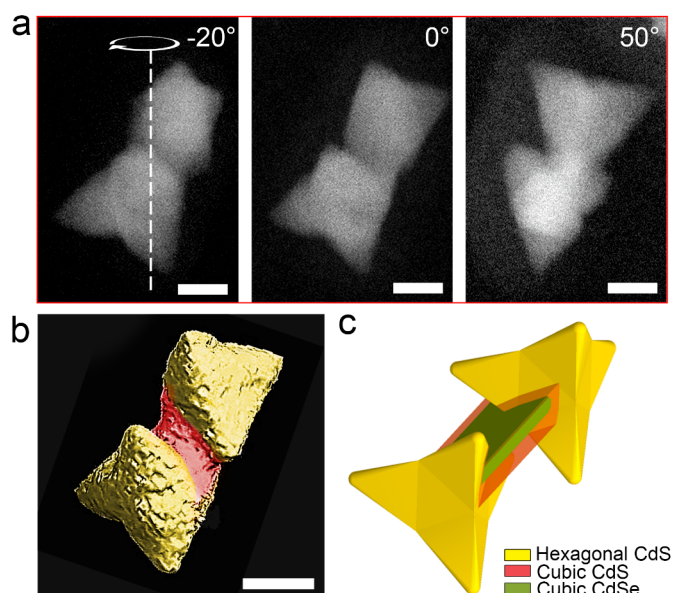

**Figure 2. a,** Selection of HAADF-STEM images of a NPL-i-TD, extracted from a tilt-series. **b,** Isosurface visualization of the reconstructed volume of a single NPL-i-TD particle, as obtained by HAADF-STEM tomography. The isosurface has been manually coloured in order to evidence the different components of the particle: the two CdS tetrahedron-shaped regions (with hexagonal crystal structure, in yellow) develop from the CdSe NPL core, covered with a cubic CdS shell, coloured in red (with cubic crystal structure). Scale bars: 10 nm. **c,** Sketch of a NPL-i-TD NC highlighting the different parts of the synthesized heterostructures.

In the synthesis of branched core/shell NCs accurate control over the injection temperature, concentration of seeds, and the use of $CdCl_2$ in the reaction are paramount to gain control over the particle shape.[27] In the case of the NPLs-i-TDs, we have found that by varying the amount of seeds used in the reaction, we can tune the ratio of the core to shell material. A lower amount of seeds (2 nmol) leads to more developed tetrahedra in highly monodisperse samples (Figure S7) while a drastic reduction in the seed content (0.4 nmol) yields NCs with different shapes and multiple pods (see Figure S8). The presence of larger pods is reflected in the XRD patterns, with a stronger intensity of the diffraction peaks associated to hexagonal CdS (Figure S9). The increase in the intensity of Bragg peaks related to the hexagonal CdS phase in the XRD pattern is due to the increase in the volume of the tetrahedra when decreasing the amount of CdSe seeds in the synthesis. This is a further confirmation of the hexagonal CdS crystal structure of the tetrahedra, assessed by HRTEM, which is in agreement with other branched Cd-chalcogenides heterostructures.[8, 22, 28] Moreover, the volume of the CdS shell in the NPL-i-TDs is much larger than that of the core, as demonstrated by TEM analysis (see Table S2). In particles synthesized by using 2 nmol of CdSe seeds, the volume of the CdSe core is 310 nm$^3$, while that of the surrounding CdS shell with cubic lattice is roughly 1200 nm$^3$, and the overall volume of the hexagonal lattice pods is around 1900 nm$^3$. For a quasi-type-II heterostructure configuration, this provides a very large volume in which the electrons can be delocalized, while the holes should be confined in the CdSe core.

For the optical characterisation we focus on the largest NPL-i-TDs obtained from 2 nmol seed concentration. First, we discuss their optical properties starting from the seeds. The CdSe NPLs manifest a strong quantum confinement, which leads to a band gap of 2.45 eV (first absorbance peak in Figure 3a), and in slightly Stokes-shifted green emission at 510 nm (2.43 eV). The core/shell CdSe/CdS NPL-i-TDs show two absorbance peaks at 650 nm and 600 nm, and a marked increase at around 500 nm (Figure 3b) that is related to absorption from the CdS shell. Here, the double peak observed in the absorption spectrum from the CdSe cores in Figure 3a is translated to the NPL-i-TDs, and therefore the two peaks in Figure 3b should also originate from transitions related to heavy and light/split-off hole levels in the CdSe core.[5e] The emission of the NPL-i-TDs is at 657 nm, and therefore strongly red-shifted by almost 150 nm with respect to the CdSe NPL seeds.

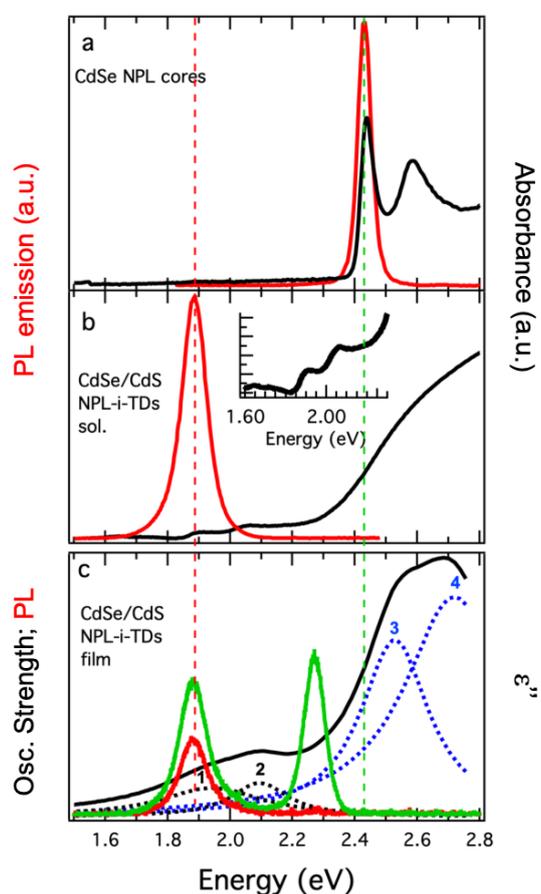

**Figure 3.** (**a,b**) Absorption and emission spectra of the CdSe NPL cores (**a**) and of the CdSe/CdS NPL-i-TDs (**b**) recorded in hexane and toluene, respectively. The inset in (**b**) shows a magnified view of the region around 2 eV in the absorption spectrum. (**c**) Imaginary part of the dielectric permittivity of a NPL-i-TDs film (full black line) and the oscillator strength of the damped oscillators (dashed lines), plotted together with the film emission under low (3.4 mW in red line) and high (4.8 mW in green line) excitation power.

Such a large red-shift in the emission of NPLs induced by a thick CdS shell has already been reported for core/shell NPLs,[24] and it was attributed to electron delocalization. Differences in the volume of the anisotropic CdS shell (however remaining in the regime with at least 1.5 nm shell thickness) do lead to only minor shifts of the NPL-i-TD emission of about 2 nm (from 659 nm to 657 nm; the FWHM increases from 28 nm to 33 nm with increasing tetrahedra size;

see Tables S2, S3 and Figure S10). The PLQY of the NPL-i-TDs in solution is around 42 %, and this value is retained for NPL-i-TD films. The absorption and emission spectra of a film of NPL-i-TDs are reported in Figure 3c.

To assess the energy levels of the optical transitions we performed spectroscopic ellipsometry and measured the complex dielectric permittivity of the NPL-i-TD films. From the ellipsometrical parameters $\psi$ and $\Delta$, a pseudo-dielectric function for the NC film can be obtained and fitted with a series of damped harmonic oscillators.[29] The complex dielectric permittivity can then be expressed as:

$$\tilde{\varepsilon} = \sum_{j=1}^{N} \frac{\alpha_j E_{0,j}}{E_{0,j}^2 - E^2 - i\beta_j E} \quad (1)$$

Here $E_{0,j}$ is the central frequency (band-gap), $\beta_j$ the FWHM, $\alpha_j$ the oscillator strength of the $j^{th}$ oscillator describing a particular energy transition, and $E$ is the energy. We obtained good fitting with four oscillator terms, and the parameters associated to the four transitions are reported in Table 1.

**Table 1.** Energy ($E_0$), oscillator strength ($\alpha$), and FWHM ($\beta$) of the four peaks in Figure 3c determined by spectroscopic ellipsometry.

|        | $E_0$ (eV) | $\alpha$ | $\beta$ |
|--------|-----------|----------|---------|
| Peak 1 | 1.95      | 0.044    | 0.401   |
| Peak 2 | 2.10      | 0.028    | 0.221   |
| Peak 3 | 2.53      | 0.179    | 0.266   |
| Peak 4 | 2.73      | 0.326    | 0.388   |

To consolidate the association of the oscillators to the optical transitions of the NPL-i-TDs, we compared the fitted peaks in Figure 3c with the experimental absorbance spectrum in Figure 3b and found very good agreement. The emission of the NPL-i-TD films shows a single peak at low excitation fluence centred at 661 nm, which is slightly red-shifted with respect to the emission from solution in Figure 3b. Interestingly, we observe a second emission peak in the green, at 544 nm (2.28 eV), from the sample with the largest CdS pods (prepared by using 2.0 nmol of CdSe seeds in the synthesis) under excitation at high pump fluence (>4.8 mW).

This behaviour is analysed in more detail in Figure 4, where the emission spectra at different pump power are plotted. The (red) PL peak at 1.88 eV first increases approximately linearly with pump power, and then its intensity saturates around 4 mW. At approximately the same pump power the second (green) peak, at 2.28 eV, appears and gains quickly in intensity. Consequently, the NPL-i-TDs film manifests dual emission with well separated emission peaks in the red and green part of the visible spectrum at excitation power higher than 4.8 mW.

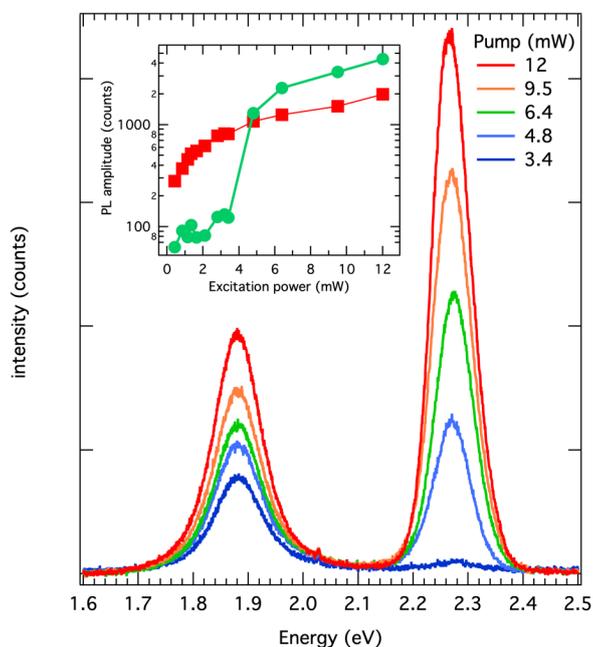

**Figure 4.** PL spectra of a drop-cast film of NPL-i-TDs excited with femtosecond pulsed laser at 400 nm at different excitation power. The inset shows the PL amplitude as a function of the excitation power for the peaks at 1.88 eV (red) and 2.28 eV (green).

We could obtain further insight into the band structure of the core/shell NPL-i-TDs through ultraviolet photoelectron spectroscopy (UPS).[10b] Figure 5a shows the full UPS spectrum versus the binding energy, where the Fermi level is set at 0 eV.

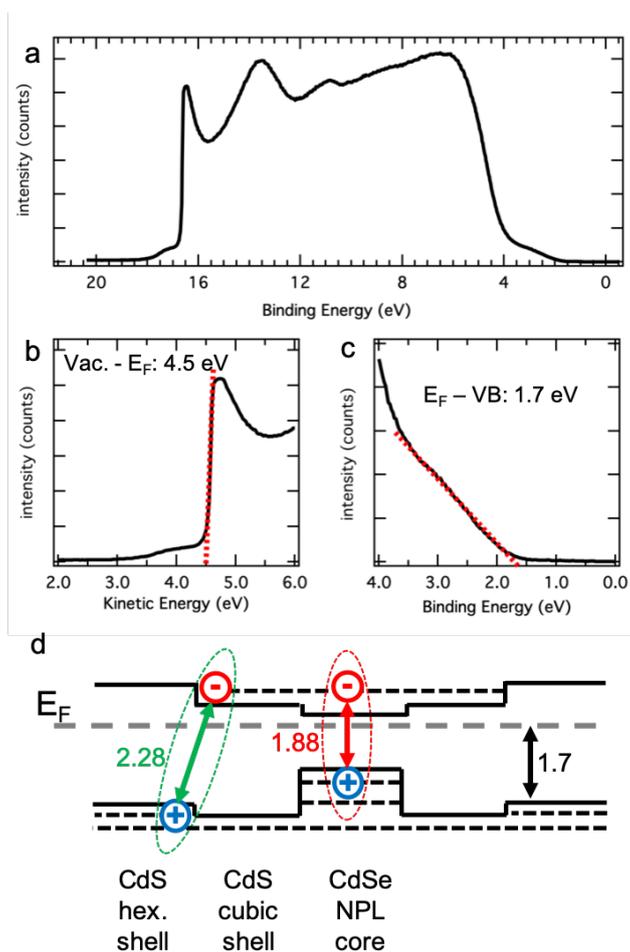

**Figure 5. a-c**, Ultraviolet photoelectron spectrum recorded from the NPL-i-TDs sample that allows to identify the valence band and Fermi level energies with respect to vacuum. **d,** Flat band diagram of the NPL-i-TDs with CdSe core, and cubic and hexagonal CdS shell. The numbers give the energy differences in eV, and the dashed ellipses indicate the two emitting channels of the NPL-i-TDs.

The low binding energy onset of 1.7 eV (Figure 5b) corresponds to the difference between the Fermi level and the top of valence band, while the signal onset at the high binding energy (low kinetic energy) side (Figure 5c) yields the energy difference from the Fermi level to the vacuum level. Combining these values with the information from the optical transitions (hexagonal CdS band gap of 2.53 eV, emission at 1.88 eV and 2.28 eV) we sketch the band alignment of the core/shell heterostructure as depicted in Figure 5d. Here we assigned the low binding energy offset to the hexagonal CdS valence band. Furthermore, we assumed a *quasi* type-II charge distribution for the CdSe/CdS system, as reported in literature for CdSe cores with strong confinement.[30] For the cubic and hexagonal structure of CdS, we considered a conduction band offset of ca. 115 meV and valence band offset of ca 50 meV,[31] which could in the NPL-i-TDs slightly deviate due to strain or confinement effects. Therefore, the band gap of the hexagonal CdS shell is slightly larger and is shifted upwards in energy with respect to the cubic one. The green emission that we observe from the NPL-i-TDs is at lower energy than that, for example, reported from CdSe/CdS dot-in-rods and tetrapods at high pump fluency.[32] Considering the band diagram in Figure 5d, it is plausible that this lower emission energy in the NPL-i-TDs results from the type-II offset between the cubic and hexagonal CdS phase. Such radiative recombination of the excitons in the shell at higher pump fluence occurs when the core recombination channel is saturated, which is corroborated by the fluence threshold for the observation of the green emission. Furthermore, we only observe the dual emission from the NPL-i-TD sample with the thickest shell (2 nmol seed concentration).

Figure 6a shows the emission spectra of a drop-casted NPL-i-TDs film under femtosecond pulsed excitation ($\lambda = 400$ nm) and increasing pump fluence. We observe an ASE band at the high-energy side of the PL peak, which indicates repulsive exciton interactions.[33] The ASE threshold is 270 $\mu$J/cm$^2$, which is significantly higher that what is obtained on CdSe core-only[34] and on CdSe/CdS NPLs with a thin shell,[5c] but within one order of magnitude comparable to other CdSe/CdS core/shell structures.[18a, 23b, 32a, 33] While the complex shell morphology of the NPLs-i-TDs is not optimal for ASE and lasing, these structures provide a peculiar exciton delocalization, and therefore, they can be very appealing for applications where electron and holes should be localized in different portions of the nanocrystal volume.

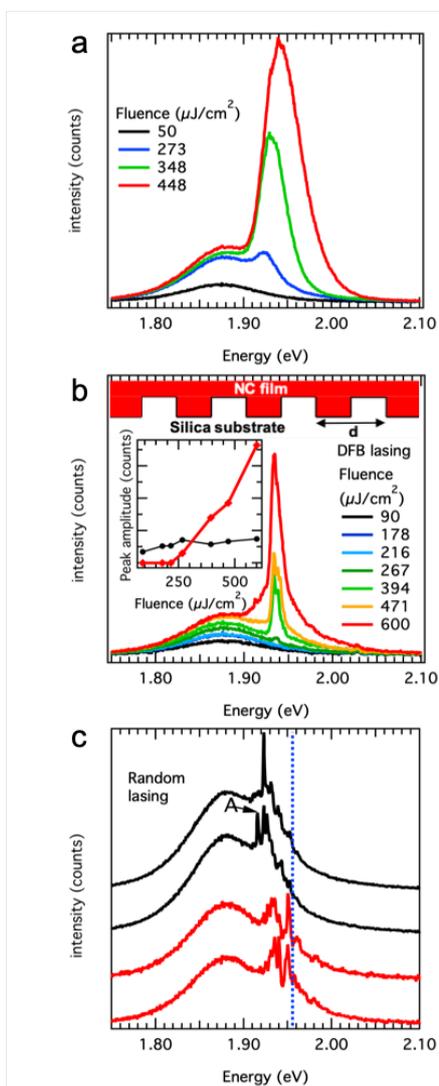

**Figure 6. a**, Emission recorded from a drop-casted film of NPL-i-TDs that manifest an ASE threshold of 270μJ/cm$^2$. **b**, Distributed feedback (DFB) lasing from NPL-i-TDs drop-casted on a silica grating with a periodicity of d = 350 nm. A lasing peak at 1.93 eV is observed. **c,** Random lasing from a drop-casted NPL-i-TDs film on a planar substrate for two different pump fluences, ∼450 μJ/cm$^2$ (black) and ∼350 μJ/cm$^2$ (red). Here the fluctuating appearance of the lasing peaks is a characteristic signature of random lasing. The peak labelled with "A", for example, appears only in the spectra recorded with higher fluence. The spectral range in which the random lasing peaks occur shifts to lower energy with increasing pump fluence. The dashed blue line indicates the energy of the lasing peak in the DFB geometry. The spectra were offset vertically for clarity.

The FWHM of the ASE peak increases with increasing pump fluence, with the peak broadening occurring on the high energy side. To test the lasing capabilities of our samples, we chose distributed feedback (DFB) resonators as a cavity since these are technologically highly appealing for solution processable materials. Here the lasing device can be simply fabricated by drop-casting the emitter solution on a silica substrate into which a suitable grating was etched, thus filling the structure with NCs. For a DFB the cavity mode is determined by the Bragg-Snell law:[35]

$$m \lambda = 2 d \sqrt{n_{eff}^2 - \sin^2 \Theta} \qquad (2)$$

where $d$ is the grating periodicity, $n_{eff}$ is the effective refractive index of the emitting film, $m$ is an integer, and $\Theta$ is the angle under which the emission is detected. We obtained lasing with a grating periodicity $d = 350$ nm under a detection angle of 5° with respect to the surface normal. Combining these parameters and setting $m = 2$ yields to an effective refractive index of $n_{eff} = 1.83$. This value is in excellent agreement with the one that we obtained by ellipsometry from NPL-i-TD films ($n_{eff} = 1.86$).

Interestingly, drop-cast films of NPL-i-TDs on planar glass substrates also show random lasing, as demonstrated in Figure 6c and S11. Here a set of sharp peaks (with ca. 0.5 nm FWHM) in emission is observed that fluctuate with time, as evident in the two consecutive spectra for two different pump fluences reported in Figure 6c (in red and black, respectively). The statistical appearance of the lasing peaks is a characteristic feature of random lasing, as for example evident for the peak labelled 'A' that only appears in the second spectrum recorded at higher fluence.

**Conclusions**

We showed that the seeded-growth protocol for fabricating branched nanocrystals, such as CdSe/CdS octapods, can be applied to highly anisotropic seeds, in particular to CdSe nanoplatelets with only few monolayers thickness. The resulting nanocrystals maintain the CdSe nanoplatelet core, and the overall length of the platelets, but manifest a double CdS shell composed of an inner part with cubic crystal structure and an outer one, made of two tetrahedra with hexagonal crystal structure. The shell increases in thickness towards the short edges of the nanoplatelets, forming overall bone-shaped nanocrystals. The thick shell leads to an excellent stabilization of the optical properties in thin films, and results in a type-II band alignment that favours shell-related transitions at high excitation fluence. The bone-shape CdSe/CdS core/shell heterostructures show high and stable PLQY for red emission from the band edge, and dual emission in the red and green spectral region under increased pump fluency, as well as ASE and lasing from thin films. Overcoating such structures with an additional ZnS shell could be a promising strategy to further increase their luminous efficiency. The design of complex shell morphologies that consist of different lattice geometries, and which manifest gradients in thickness, opens new perspectives for core/shell heterostructure materials. Different shell thickness morphologies could be employed to control the multiexciton distribution in such heterostructures, and could lead to dipole moments that depend on the exciton population and excitation fluence.

**Experimental Section**

**Materials.** Cadmium chloride ($CdCl_2$, 99.99 %) and Octadecylphosphonic acid (ODPA, 97 %) were purchased from Sigma-Aldrich; cadmium oxide powder (CdO, 99.999 %), sulphur powder (S, 99 %), tri-n-octylphosphine oxide (TOPO, 99 %) and tri-n-octylphosphine (TOP, 97 %) were purchased from Strem Chemicals; hexylphosphonic acid (HPA) was purchased from Polycarbon Industries. All chemicals were used as received and all the syntheses were carried out using a standard Schlenk line.

**Synthesis of CdSe/CdS NPL-i-TDs NCs.** CdSe NPLs were synthesized according to a previously published procedure.[5f] In a three-neck flask 60 mg of CdO, 6 mg of $CdCl_2$, 3 g of TOPO, 290 mg of ODPA, and 80 mg of HPA were degassed for 1h at 140°C under stirring. The solution was heated up to 350°C under $N_2$ flow. When the solution became transparent (≈ 270°C), 2.5 mL of anhydrous TOP were injected in the flask. Once the temperature reached 350°C, the CdSe seeds (in hexane), 500 μl of TOP and 620 μL of S precursor (32 mg/ml solution in TOP) were quickly injected. The reaction was run for 10 minutes before quickly cooling it down to room temperature. When the temperature was below 100°C, 3 mL of anhydrous toluene was added to the resulting product. The particles were then washed twice via a solvent/antisolvent protocol with an anhydrous mixture of ethanol and methanol and re-dispersed in toluene. The size and shape of the NPL-i-TDs were tuned using different amounts of CdSe seeds that were synthesized following a reported protocol.[5f]

**Structural, optical, and surface characterization.** Bright-field transmission electron microscopy (BF-TEM) analyses were carried out on a JEM-1011 (JEOL, Tokyo-Japan) microscope, operated at 100 kV acceleration voltage and equipped with a tungsten thermionic electron source. HAADF-STEM images and tilted series (from −74° to +74°, in steps of 2° at high angles and 5° between −30° and +30°) were acquired using a FEI Tecnai $G^2$ F20 TEM, operated at 200 kV. Fiducial-less alignment routine from IMOD[36] was used for alignment and the volume was reconstructed by back projection using the Tomoj plugin of ImageJ.[37] UCSF Chimera software was used to perform the volume rendering.[38] High-resolution TEM (HRTEM) and STEM-EDS elemental composition analysis were carried out with an image-Cs-corrected JEM-2200FS instrument (JEOL), equipped with a Bruker 60 mm$^2$ XFlash 5060 silicon drift detector. For elemental maps, the Kα peaks were used for S and Se, and the Lα peak was used for Cd. For all TEM analyses the samples were prepared by drop-casting (3-5 drops, 5 μL each) onto ultrathin carbon-coated Cu grids, except for HRTEM analyses, which needed ultrathin-on-holey-carbon-coated Cu grids.

Powder X-Ray Diffraction patterns were collected on a PANalytical Empyrean X-ray diffractometer equipped with a 1.8 kW CuKα ceramic X-ray tube, PIXcel3D 2x2 area detector and operating at 45 kV and 40 mA. The NC suspensions were drop-cast on a zero-diffraction Si substrate for XRD analysis.

The absorption spectra of NPL-i-TDs were collected from toluene suspensions by using a Varian Cary 5000 ultraviolet–visible–near infrared (UV–vis–NIR) spectrophotometer. The PL spectra were recorded with a Horiba FluoroMax 4 spectrometer, exciting at 340 nm and filtering the emitted light with a low-pass filter at 370 nm. Edinburgh Instruments fluorescence spectrometer (FLS920) equipped with a calibrated integrating sphere and a Xenon lamp with monochromator was used to carry out the PLQY measurements of the NCs dispersed in toluene, exciting them at 450 nm. The ASE and laser emission was measured using a Ti:sapphire laser (Coherent Legend Elite seeded by a Ti:sapphire fs laser, λ = 405 nm, 70 fs pulse FWHM and repetition rate of 1 kHz) for excitation, and an Ocean Optics HR4000 spectrometer coupled to an optical fiber for collection. The ASE measurements were performed by focusing the excitation beam with a cylindrical lens onto the sample, thus obtaining a stripe-shaped beam profile, collecting at ∼90° with respect to the excitation beam. Laser emission was recorded with a spherical lens at ∼5° with respect to the excitation beam, using a long-pass filter ($\lambda_{cut-off}$ = 450 nm) to prevent the laser beam from reaching the detector. The custom-made DFB gratings on a silica substrate were purchased from NIL Technology.

The spectroscopic ellipsometry characterization was performed with a Vertical Vase ellipsometer by Woollam, in the range from 300-900 nm. Spectroscopic analysis was done at three different angles (50°, 60° and 70°) with a step of 3 nm. The resolution of recorded spectra is 3 nm, and all spectra have been normalized to the intensity of the Xe lamp. ASE, random lasing, and PL under different pump fluency measurements were recorded from drop-casted films prepared by depositing 100 μl of a diluted suspension (200 μl of toluene were added to 100 μl of the washed nanocrystal suspension) of nanocrystals into the different substrates of 15 x 15 mm.

Ultraviolet Photoelectron Spectroscopy (UPS) analysis was performed on drop-cast films (prepared by depositing 5 μl of the washed nanocrystal suspension into 30 nm Au-coated Si substrates of 5 mm x 5 mm) to estimate the position of the valence band maximum (VBM) and work function of the material under investigation. The measurements were carried out with a Kratos Axis Ultra$^{DLD}$ spectrometer, using a He I (21.22 eV) discharge lamp, on an area of 55 μm in diameter, at a pass energy of 5 eV and with a dwell time of 100 ms. The work function (that is, the position of the Fermi level with respect to the vacuum level) was measured from the threshold energy for the emission of secondary electrons during He I excitation. A −9.0 V bias was applied to the sample to precisely determine the low-kinetic-energy cutoff, as discussed by Helander and colleagues.[39] Then, the position of the VBM versus the vacuum level was estimated by measuring its distance from the Fermi level.[40] In an attempt to evidence the presence of allowing at the core/shell interface we collected a Raman spectrum from NPLs-i-TDs deposited by drop-casting on Au-coated glass. A Renishaw micro-Raman equipped with a 50× (0.75 N.A.) objective with an excitation wavelength of 514 nm was used. The incident power was kept below 1 mW to avoid damage of the samples during the measurement with 5 min of accumulation to minimize the noise.

[Further details of the crystal structure investigations may be obtained from the Fachinformationszentrum Karlsruhe, 76344 Eggenstein-Leopoldshafen (Germany), on quoting the depository number ICSD-81925, ICSD-154186, ICDD 96-101-1055, ICDD 96-900-0109, and ICDD 98-018-6011].

WILEY-VCH

**Supporting Information**
Supporting Information is available from the Wiley Online Library or from the author.

**Acknowledgements**
MA and RK acknowledge financial support by the EU Horizon2020 MSCA Rise project "COMPASS-691185". This project has received funding from the European Research Council (ERC) under the European Union's Horizon 2020 research and innovation programme (grant agreement No. 714876 PHOCONA).

**Conflict of interests**
The authors declare no conflict of interest.

Received: ((will be filled in by the editorial staff))
Revised: ((will be filled in by the editorial staff))
Published online: ((will be filled in by the editorial staff))

**References**

[1] a) D. V. Talapin, J. S. Lee, M. V. Kovalenko, E. V. Shevchenko, *Chem. Rev.* **2010**, 110, 389; b) F. Fan, O. Voznyy, R. P. Sabatini, K. T. Bicanic, M. M. Adachi, J. R. McBride, K. R. Reid, Y.-S. Park, X. Li, A. Jain, R. Quintero-Bermudez, M. Saravanapavanantham, M. Liu, M. Korkusinski, P. Hawrylak, V. I. Klimov, S. J. Rosenthal, S. Hoogland, E. H. Sargent, *Nature* **2017**, 544, 75; c) C. Dang, J. Lee, C. Breen, J. S. Steckel, S. Coe-Sullivan, A. Nurmikko, *Nat. Nanotechnol.* **2012**, 7, 335.

[2] a) W. Cho, S. Kim, I. Coropceanu, V. Srivastava, B. T. Diroll, A. Hazarika, I. Fedin, G. Galli, R. D. Schaller, D. V. Talapin, *Chem. Mater.* **2018**, 30, 6957; b) K. E. Knowles, K. H. Hartstein, T. B. Kilburn, A. Marchioro, H. D. Nelson, P. J. Whitham, D. R. Gamelin, *Chem. Rev.* **2016**, 116, 10820.

[3] a) J. Owen, L. Brus, *J. Am. Chem. Soc.* **2017**, 139, 10939; b) C. B. Murray, D. J. Norris, M. G. Bawendi, *J. Am. Chem. Soc.* **1993**, 115, 8706.

[4] X. G. Peng, L. Manna, W. D. Yang, J. Wickham, E. Scher, A. Kadavanich, A. P. Alivisatos, *Nature* **2000**, 404, 59.

[5] a) S. Ithurria, M. D. Tessier, B. Mahler, R. P. S. M. Lobo, B. Dubertret, A. L. Efros, *Nat. Mater.* **2011**, 10, 936; b) E. Lhuillier, S. Pedetti, S. Ithurria, B. Nadal, H. Heuclin, B. Dubertret, *Acc. Chem. Res.* **2015**, 48, 22; c) C. She, I. Fedin, D. S. Dolzhnikov, A. Demortière, R. D. Schaller, M. Pelton, D. V. Talapin, *Nano Lett.* **2014**, 14, 2772; d) V. I. Klimov, A. A. Mikhailovsky, D. W. McBranch, C. A. Leatherdale, M. G. Bawendi, *Science* **2000**, 287, 1011; e) S. Christodoulou, J. I. Climente, J. Planelles, R. Brescia, M. Prato, B. Martín-García, A. H. Khan, I. Moreels, *Nano Lett.* **2018**, 18, 6248; f) G. H. V. Bertrand, A. Polovitsyn, S. Christodoulou, A. H. Khan, I. Moreels, *Chem. Commun.* **2016**, 52, 11975.

[6] a) X. G. Peng, M. C. Schlamp, A. V. Kadavanich, A. P. Alivisatos, *J. Am. Chem. Soc.* **1997**, 119, 7019; b) M. A. Hines, P. Guyot-Sionnest, *J. Phys. Chem.* **1996**, 100, 468.

[7] a) W. Shi, H. Zeng, Y. Sahoo, T. Y. Ohulchanskyy, Y. Ding, Z. L. Wang, M. Swihart, P. N. Prasad, *Nano Lett.* **2006**, 6, 875; b) A. Vaneski, A. S. Susha, J. Rodríguez-Fernández, M. Berr, F. Jäckel, J. Feldmann, A. L. Rogach, *Adv. Funct. Mater.* **2011**, 21, 1547.

[8] L. Manna, D. J. Milliron, A. Meisel, E. C. Scher, A. P. Alivisatos, *Nat. Mater.* **2003**, 2, 382.

[9] a) T. Li, B. Xue, B. Wang, G. Guo, D. Han, Y. Yan, A. Dong, *J. Am. Chem. Soc.* **2017**, 139, 12133; b) C. Y. Toe, Z. Zheng, H. Wu, J. Scott, R. Amal, Y. H. Ng, *J. Phys. Chem. C* **2018**, 122, 14072.

[10] a) M. Liu, O. Voznyy, R. Sabatini, F. P. García de Arquer, R. Munir, A. H. Balawi, X. Lan, F. Fan, G. Walters, A. R. Kirmani, S. Hoogland, F. Laquai, A. Amassian, E. H. Sargent, *Nat. Mater.* **2016**, 16, 258; b) P. Rastogi, F. Palazon, M. Prato, F. Di Stasio, R. Krahne, *ACS Appl. Mater. Interfaces* **2018**, 10, 5665; c) G. González-Rubio, J. González-Izquierdo, L. Bañares, G. Tardajos, A. Rivera, T. Altantzis, S. Bals, O. Peña-Rodríguez, A. Guerrero-Martínez, L. M. Liz-Marzán, *Nano Lett.* **2015**, 15, 8282; d) L. K. Bogart, G. Pourroy, C. J. Murphy, V. Puntes, T. Pellegrino, D. Rosenblum, D. Peer, R. Lévy, *ACS Nano* **2014**, 8, 3107.

[11] a) T. D.V., R. A.L., K. A., W. Haase M., H., *Nano Lett.* **2001**, 1, 207; b) Y. Chen, J. Vela, H. Htoon, J. L. Casson, D. J. Werder, D. A. Bussian, V. I. Klimov, J. A. Hollingsworth, *J. Am. Chem. Soc.* **2008**, 130, 5026.

[12] S. Brovelli, R. D. Schaller, S. A. Crooker, F. García-Santamaría, Y. Chen, R. Viswanatha, J. A. Hollingsworth, H. Htoon, V. I. Klimov, *Nat. Commun.* **2011**, 2, 280.

[13] S. Christodoulou, F. Rajadell, A. Casu, G. Vaccaro, J. Q. Grim, A. Genovese, L. Manna, J. I. Climente, F. Meinardi, G. Rainò, T. Stöferle, R. F. Mahrt, J. Planelles, S. Brovelli, I. Moreels, *Nat. Commun.* **2015**, 6, 7905.

[14] a) L. Carbone, C. Nobile, M. De Giorgi, F. D. Sala, G. Morello, P. Pompa, M. Hytch, E. Snoeck, A. Fiore, I. R. Franchini, M. Nadasan, A. F. Silvestre, L. Chiodo, S. Kudera, R. Cingolani, R. Krahne, L. Manna, *Nano Lett.* **2007**, 7, 2942; b) E. Bladt, R. J. A. van Dijk-Moes, J. Peters, F.

Montanarella, C. de Mello Donega, D. Vanmaekelbergh, S. Bals, *J. Am. Chem. Soc.* **2016**, 138, 14288.

[15] a) D. Dorfs, T. Franzl, R. Osovsky, M. Brumer, E. Lifshitz, T. A. Klar, A. Eychmueller, *Small* **2008**, 4, 1148; b) E. Groeneveld, C. de Mello Donega, *J. Phys. Chem. C.* **2012**, 116, 16240.

[16] a) F. García-Santamaría, S. Brovelli, R. Viswanatha, J. A. Hollingsworth, H. Htoon, S. A. Crooker, V. I. Klimov, *Nano Lett.* **2011**, 11, 687; b) F. García-Santamaría, Y. Chen, J. Vela, R. D. Schaller, J. A. Hollingsworth, V. I. Klimov, *Nano Lett.* **2009**, 9, 3482; c) M. Nasilowski, P. Spinicelli, G. Patriarche, B. Dubertret, *Nano Lett.* **2015**, 15, 3953; d) J. I. Climente, J. L. Movilla, J. Planelles, *Small* **2012**, 8, 754.

[17] Y. Kelestemur, D. Dede, K. Gungor, C. F. Usanmaz, O. Erdem, H. V. Demir, *Chem. Mater.* **2017**, 29, 4857.

[18] a) F. Di Stasio, A. Polovitsyn, I. Angeloni, I. Moreels, R. Krahne, *ACS Photonics* **2016**, 3, 2083; b) A. Polovitsyn, A. H. Khan, I. Angeloni, J. Q. Grim, J. Planelles, J. I. Climente, I. Moreels, *ACS Photonics* **2018**, 5, 4561.

[19] R. Vaxenburg, E. Lifshitz, *Phys. Rev. B* **2012**, 85, 075304.

[20] D. V. Talapin, J. H. Nelson, E. V. Shevchenko, S. Aloni, B. Sadtler, A. P. Alivisatos, *Nano Lett.* **2007**, 7, 2951.

[21] A. Fiore, R. Mastria, M. G. Lupo, G. Lanzani, C. Giannini, E. Carlino, G. Morello, M. De Giorgi, Y. Li, R. Cingolani, L. Manna, *J. Am. Chem. Soc.* **2009**, 131, 2274.

[22] S. Deka, K. Miszta, D. Dorfs, A. Genovese, G. Bertoni, L. Manna, *Nano Lett.* **2010**, 10, 3770.

[23] a) B. M. Saidzhonov, V. F. Kozlovsky, V. B. Zaytsev, R. B. Vasiliev, *J. Lumin.* **2019**, 209, 170; b) Y. Gao, M. Li, S. Delikanli, H. Zheng, B. Liu, C. Dang, T. C. Sum, H. V. Demir, *Nanoscale* **2018**, 10, 9466; c) A. Polovitsyn, Z. Dang, J. L. Movilla, B. Martín-García, A. H. Khan, G. H. V. Bertrand, R. Brescia, I. Moreels, *Chem. Mater.* **2017**, 29, 5671.

[24] A. A. Rossinelli, A. Riedinger, P. Marqués-Gallego, P. N. Knüsel, F. V. Antolinez, D. J. Norris, *Chem. Commun.* **2017**, 53, 9938.

[25] R. Brescia, K. Miszta, D. Dorfs, L. Manna, G. Bertoni, *J. Phys. Chem. C.* **2011**, 115, 20128.

[26] a) G. Bertoni, V. Grillo, R. Brescia, X. Ke, S. Bals, A. Catellani, H. Li, L. Manna, *ACS Nano* **2012**, 6, 6453; b) S. Ghosh, R. Gaspari, G. Bertoni, M. C. Spadaro, M. Prato, S. Turner, A. Cavalli, L. Manna, R. Brescia, *ACS Nano* **2015**, 9, 8537.

[27] a) M. R. Kim, K. Miszta, M. Povia, R. Brescia, S. Christodoulou, M. Prato, S. Marras, L. Manna, *ACS Nano* **2012**, 6, 11088; b) A. Castelli, J. de Graaf, M. Prato, L. Manna, M. P. Arciniegas, *ACS Nano* **2016**, 10, 4345.

[28] D. J. Milliron, S. M. Hughes, Y. Cui, L. Manna, J. Li, L.-W. Wang, A. P. Alivisatos, *Nature* **2004**, 430, 190.

[29] a) G. E. J. Jr., F. A. Modine, *Appl. Phys. Lett.* **1996**, 69, 371; b) M. Palei, V. Caligiuri, S. Kudera, R. Krahne, *ACS Appl. Mater. Interfaces* **2018**, 10, 22356.

[30] a) A. Sitt, F. D. Sala, G. Menagen, U. Banin, *Nano Lett.* **2009**, 9, 3470; b) G. Rainò, T. Stöferle, I. Moreels, R. Gomes, J. S. Kamal, Z. Hens, R. F. Mahrt, *ACS Nano* **2011**, 5, 4031; c) L. Wang, K. Nonaka, T. Okuhata, T. Katayama, N. Tamai, *J. Phys. Chem. C.* **2018**, 122, 12038.

[31] E. United States. Department of, L. National Renewable Energy, S. United States. Department of Energy. Office of, I. Technical, *ZB/WZ Band Offsets and Carrier Localization in CdTe Solar Cells*, United States. Department of Energy, **2000**.

[32] a) R. Krahne, M. Zavelani-Rossi, M. G. Lupo, L. Manna, G. Lanzani, *Appl. Phys. Lett.* **2011**, 98, 063105; b) A. A. Lutich, C. Mauser, E. Da Como, J. Huang, A. Vaneski, D. V. Talapin, A. L. Rogach, J. Feldmann, *Nano Lett.* **2010**, 10, 4646.

[33] C. Gollner, J. Ziegler, L. Protesescu, D. N. Dirin, R. T. Lechner, G. Fritz-Popovski, M. Sytnyk, S. Yakunin, S. Rotter, A. A. Yousefi Amin, C. Vidal, C. Hrelescu, T. A. Klar, M. V. Kovalenko, W. Heiss, *ACS Nano* **2015**, 9, 9792.

[34] J. Q. Grim, S. Christodoulou, F. Di Stasio, R. Krahne, R. Cingolani, L. Manna, I. Moreels, *Nat. Nanotechnol.* **2014**, 9, 891.

[35] L. Berti, M. Cucini, F. Di Stasio, D. Comoretto, M. Galli, F. Marabelli, N. Manfredi, C. Marinzi, A. Abbotto, *J. Phys. Chem. C.* **2010**, 114, 2403.
[36] J. R. Kremer, D. N. Mastronarde, J. R. McIntosh, *Journal of Structural Biology* **1996**, 116, 71.
[37] a) C. A. Schneider, W. S. Rasband, K. W. Eliceiri, *Nat. Met.* **2012**, 9, 671; b) C. Messaoudil, T. Boudier, C. O. S. Sorzano, S. Marco, *BMC Bioinform.* **2007**, 8, 288.
[38] E. F. Pettersen, T. D. Goddard, C. C. Huang, G. S. Couch, D. M. Greenblatt, E. C. Meng, T. E. Ferrin, *J. Comput. Chem.* **2004**, 25, 1605.
[39] M. G. Helander, M. T. Greiner, Z. B. Wang, Z. H. Lu, *Appl. Surf. Sci.* **2010**, 256, 2602.
[40] A. Calloni, A. Abate, G. Bussetti, G. Berti, R. Yivlialin, F. Ciccacci, L. Duò, *J. Phys. Chem. C.* **2015**, 119, 21329.

The table of contents entry:

**A new type of highly anisotropic core/shell nanocrystals is developed** with a shell made of different crystal lattices and thickness, and a core that preserve its nanoplatelet geometry despite the high synthesis temperatures. By exploiting such complex shape, double emission, amplified spontaneous emission, random lasing, and distributed feedback lasing are achieved, opening new views on engineering optical properties through exotic particle architectures.

**Keyword**
Core/Shell Branched Nanoemitters.

A. Castelli, B. Dhanabalan, A. Polovitsyn, V. C Caligiuri, F. Di Stasio, A. Scarpelini, R. Brescia, M. Palei, B. Martin-Garcia, M. Prato, L. Manna, I. Moreels, R. Krahne*, and M. P. Arciniegas*

**Core/Shell CdSe/CdS Bone-Shaped Nanocrystals with a Thick and Anisotropic Shell as Optical Emitters**

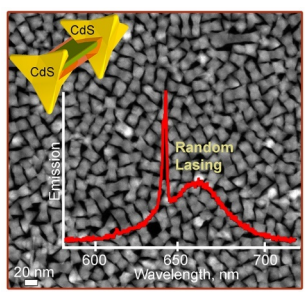